\documentclass[journal=jacsat,manuscript=article]{achemso}

\usepackage{chemformula} 
\usepackage[T1]{fontenc} 
\usepackage{comment}

\author{Elena Stellino}
\affiliation{Department of Physics and Geology, University of Perugia, via Alessandro Pascoli, Perugia, Italy}
\author{Francesca Ripanti}
\affiliation{Department of Physics, Sapienza University of Rome, P.le A. Moro 5, Rome, Italy}
\email{francesca.ripanti@uniroma1.it}
\author{Giacomo Nisini}
\affiliation{Department of Physics, Sapienza University of Rome, P.le A. Moro 5, Rome, Italy}
\author{Francesco Capitani}
\affiliation{Synchrotron SOLEIL, L’Orme des Merisiers, Saint-Aubin, Gif-sur-Yvette, France}
\author{Caterina Petrillo}
\affiliation{Department of Physics and Geology, University of Perugia, via Alessandro Pascoli, Perugia, Italy}
\author{Paolo Postorino}
\affiliation{Department of Physics, Sapienza University of Rome, P.le A. Moro 5, Rome, Italy}
\email{paolo.postorino@roma1.infn.it}

\title[An \textsf{achemso} demo]
  {Infrared study of the pressure-induced isostructural metallic transition in Mo$_{0.5}$W$_{0.5}$S$_2$}

\abbreviations{IR,NMR,UV}

\begin{document}

\begin{abstract}

Ternary compounds of Transition Metal Dichalcogenides are emerging as an interesting class of crystals with tunable electronic properties, which make them attractive for nano-electronic and optoelectronic applications. Among them, Mo$_x$W$_{1-x}$S$_2$ is one of the most studied alloys, due to the well-known, remarkable features of its binary constituents, MoS$_2$ and WS$_2$. The band-gap of this compound can be modelled varying Mo and W percentages in the sample, and its vibrational modes result from a combination of MoS$_2$ and WS$_2$ phonons. In this work, we report transmission measurements on a Mo$_{0.5}$W$_{0.5}$S$_2$ single crystal in the far-infrared range. Absorbance spectra collected at ambient conditions enabled, for the first time, a classification of the infrared-active phonons, complementary to Raman studies. High-pressure measurements allowed to study the evolution of both the lattice dynamics and the free carrier density up to 31 GPa, indicating the occurrence of an isostructural semiconductor-to-metal transition above 18 GPa.

\end{abstract}

\section{Introduction}

Among two-dimensional materials, in the last years, Transition Metal Dichalcogenides (TMDs) have proved to be one of the most promising classes of crystals in terms of both applications and fundamental studies. Semiconducting TMDs are characterized by a graphene-like layered lattice, easily exfoliable down to atomic-thick crystals \cite{Yuan2016, Manzeli2017, Peng2017}, but, at variance with graphene, their band structure exhibits finite band-gaps in the eV-scale, which are attractive for electronic devices \cite{Li2017, Wang2012, Jariwala2014}. One of the most remarkable features of TMDs is the key role of the inter-layer interaction in determining the sample properties, such as the strong relationship between the electronic structure and the number of layers. Indeed, in most semiconductors, a progressive increase of the band-gap is observed as the number of layers reduces, and an indirect-to-direct band-gap crossover arises when bilayer samples are scaled down to monolayers \cite{Yun2012, Mak2010, Zhang2013}.
\\
Due to the high heterogeneity and the outstanding structural and electronic properties of isomorphic semiconducting TMDs, the possibility to design and produce heterostructures (formed by stacking together monolayers of different crystals) and alloys (synthesized by directly mixing two different TMDs) with different band-gaps has emerged as an appealing way to tune and tailor the band-structure of these materials for nano-electronic and optoelectronic applications. In this framework, ternary compounds have been synthesized where the metal or the chalcogen contributions are adjusted with different atoms from the same element group \cite{Xie2015,Zhao2018}.
\\
Mo$_x$W$_{1-x}$S$_2$ is one of the most investigated TMD alloys, due to the vast amount of studies on its binary constituents MoS$_2$ and WS$_2$ \cite{Sahoo2013, Chakraborty2012, Mahatha_2012, Ulstrup2017, Sekine1980}. The bulk crystal shows a 2H phase and is formed by stacking together monolayer alloys via van der Waals interactions. The monolayer alloy contains one MoW plane sandwiched by two S planes, represented as  S-Mo/W-S. 
Recent work on Mo$_x$W$_{1-x}$S$_2$ has proved that the electronic band-gap at ambient conditions can be tuned by modifying the percentage of Mo and W atoms in the lattice \cite{Wang2016, Kim2017}. Raman measurements have also been performed, varying the number of layers and the stoichiometric composition, to characterize the vibrational modes of the sample \cite{Qiao2015, Zhang2015}.
\\
The application of pressure is a powerful tool to probe the inter-layer interactions in Mo$_x$W$_{1-x}$S$_2$ and to explore the tunability of the electronic and structural properties. Indeed, experimental and theoretical reports have already demonstrated that pressure can modulate the band structure of TMDs in general \cite{Stellino2020, Yang2019}, and in particular of  MoS$_2$ and WS$_2$ \cite{Nayak2014, Nayak2015}. In MoS$_2$ a semiconductor-to-metal transition has been observed at $\sim$ 19 GPa, accompanied by a structural distortion from 2H$_c$ to 2H$_a$ phase. Similarly, in WS$_2$ a metallic transition occurs at $\sim$ 22 GPa, but, at variance with the previous case, the lattice remains isomorphic. High-pressure Raman measurements on Mo$_{0.5}$W$_{0.5}$S$_2$ have suggested the absence of structural transitions up to 40 GPa \cite{Kim_2016}. However, the evolution of its electronic properties remains unknown.
\\
Here we report high-pressure transmission measurements on Mo$_{0.5}$W$_{0.5}$S$_2$ in the far-infrared (FIR) range, where simultaneous information on the lattice dynamics, complementary to Raman studies, and on the pressure-induced increase of the carrier density are accessible. This work classifies for the first time the infrared-active modes of Mo$_{0.5}$W$_{0.5}$S$_2$ at ambient conditions, analyzes their response under pressure, confirming the absence of structural transitions up to 31 GPa, and observes an abrupt increase of the carrier density above 20 GPa, associated to the onset of a semiconductor-to-metal transition.

\section{Results and discussion}
\subsection{Vibrational modes at ambient conditions}
\begin{figure}
\includegraphics[scale=0.35]{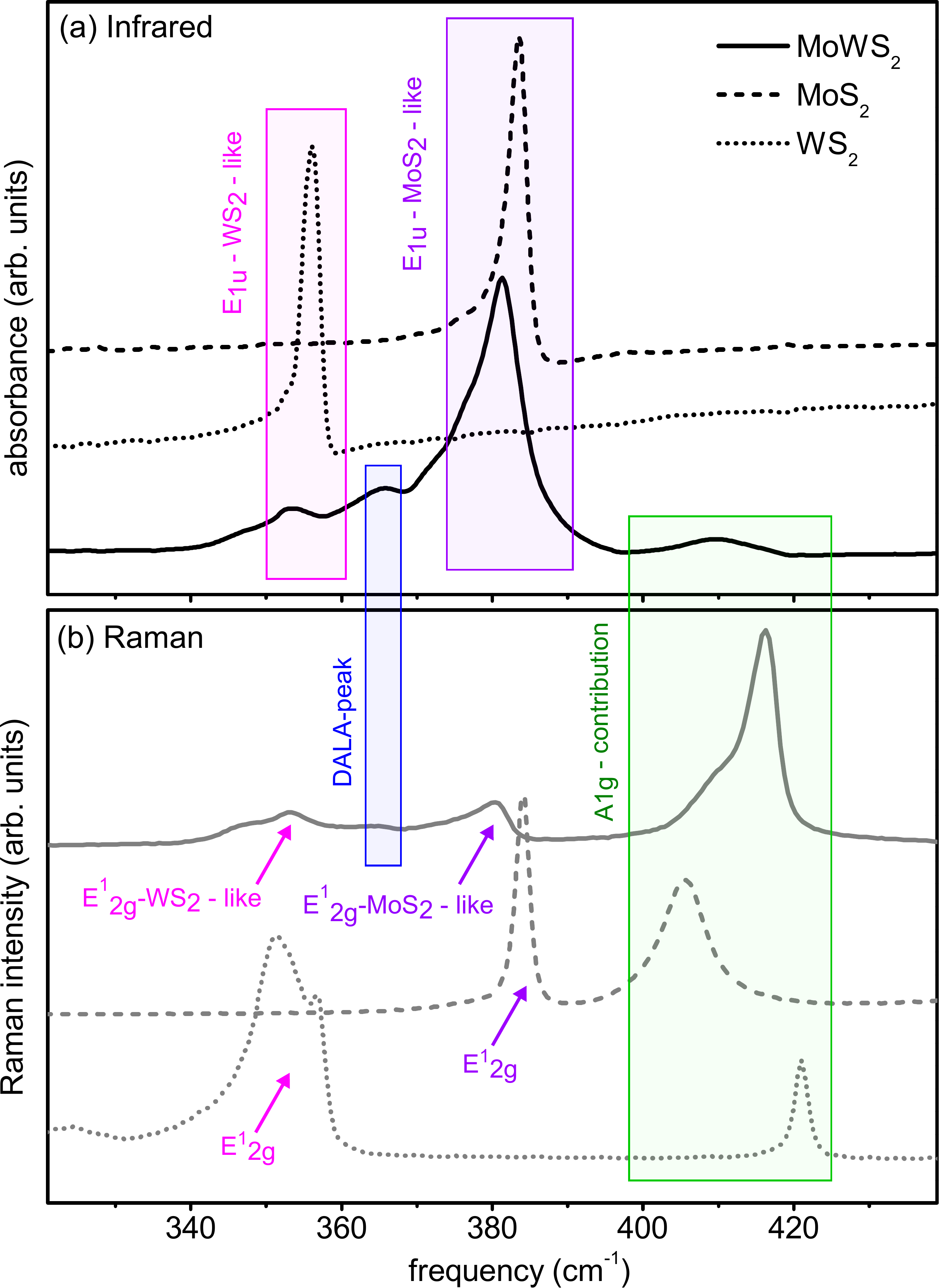}
  \caption{Infrared (a) and Raman (b) spectra of Mo$_{0.5}$W$_{0.5}$S$_2$ (continuous lines), MoS$_2$ (dashed lines), and WS$_2$ (dotted lines). Magenta and purple areas highlight the contributions of Mo$_{0.5}$W$_{0.5}$S$_2$ coming from the IR-active phonons of MoS$_2$ and WS$_2$ ($E_{1u}$). Blue and green areas highlight the contributions of Mo$_{0.5}$W$_{0.5}$S$_2$ coming from the IR-inactive, disorder-activated modes of MoS$_2$ and WS$_2$ ($A_{1g}$, DALA, see text).}
  \label{fig1}
\end{figure}

Infrared transmission measurements at ambient conditions were performed on Mo$_{0.5}$W$_{0.5}$S$_2$, WS$_2$, and MoS$_2$ crystals in the  100$\div$600 cm$^{-1}$ range. To reduce the interference effects between the sample surfaces, each crystal was exfoliated on a diamond window, thus minimizing the discontinuity of the refractive index at the substrate interface. Measurements of the background intensity, $I_0(\omega)$, were preliminarily carried out on the bare diamond. The sample transmitted intensity, $I(\omega)$, was then measured once the crystals were positioned on the diamond, allowing to determine the absorbance spectra $A(\omega)=-ln[I(\omega)/I_0(\omega)]$. Raman measurements were also collected in the same frequency range as a reference.
The comparison between absorbance and Raman spectra of Mo$_{0.5}$W$_{0.5}$S$_2$, WS$_2$, and MoS$_2$ is shown in Figure \ref{fig1}.

The absorbance spectrum of Mo$_{0.5}$W$_{0.5}$S$_2$ is characterized by the presence of four distinct bands between 300 cm$^{-1}$ and 450 cm$^{-1}$. By comparing it with the $A(\omega)$ spectra of WS$_2$ and MoS$_2$, the peaks at $\sim$ 350 cm$^{-1}$ and $\sim$ 380 cm$^{-1}$ can be assigned to the E$_{1u}$-WS$_2$-like and the E$_{1u}$-MoS$_2$-like modes respectively \cite{ONeal2016, Guo2015}. The larger bandwidth observed for the alloy peaks can be attributed to disorder-effects, which are less important in the binary crystals. It is interesting to notice that, although Mo and W atoms are present in the sample with the same percentage, the intensity of E$_{1u}$-MoS$_2$-like is far larger than that of E$_{1u}$-WS$_2$-like. We recall that the infrared phonon intensity is proportional to the square of the first derivative of the dipole moment with respect to the normal mode coordinates. The dipole moment is related to the Born effective charge tensor, i.e. the first derivative of the polarization per unit cell with respect to the atomic displacements. Based on the Density-Functional-Theory calculations reported in the literature \cite{Pike2017}, the Born charges (defined as one-third of the trace of the tensor in Cartesian coordinates) of Mo and W, for in-plane displacements, are in a 2:1 ratio, suggesting a four-times higher intensity of MoS$_2$-like mode compared with WS$_2$-like one.
The less intense peak at $\sim$ 365 cm$^{-1}$ between the two E$_{1u}$ phonons may correspond to the disorder-activated longitudinal acoustic (DALA) phonon mode \cite{Kim_2016}, while the broad band at $\sim$ 410 cm$^{-1}$ could be identified with the convoluted A$_{1g}$ modes from MoS$_2$ and WS$_2$ \cite{Zhang2015}. Interestingly, both the DALA and the A$_{1g}$-like modes, whose symmetries are not compatible with the infrared selection rules, do not appear in the binary crystals but can be activated by disorder effects in the ternary compound.

\subsection{Vibrational modes under high pressure}
Room temperature transmission measurements on a Mo$_{0.5}$W$_{0.5}$S$_2$ single crystal were performed in the  100$\div$600 cm$^{-1}$ range, on increasing pressure from 0 to 31 GPa. The background intensity, $I_0(\omega)$, was measured with the DAC filled by the hydrostatic medium (CsI) only. The sample transmitted intensity, $I(\omega)$, was then collected once the crystal was loaded into the cell, to obtain the absorbance, $A(\omega)$, at each pressure.
Interference fringes due to multiple reflections between the
diamond surfaces through the hydrostatic medium were observed in $A(\omega)$ spectra. The reduced number of oscillations and the frequency dependence of both the period and the amplitudeh of the oscillations prevent an effective subtraction in the spectra. 

\begin{figure}
\includegraphics[scale=0.55]{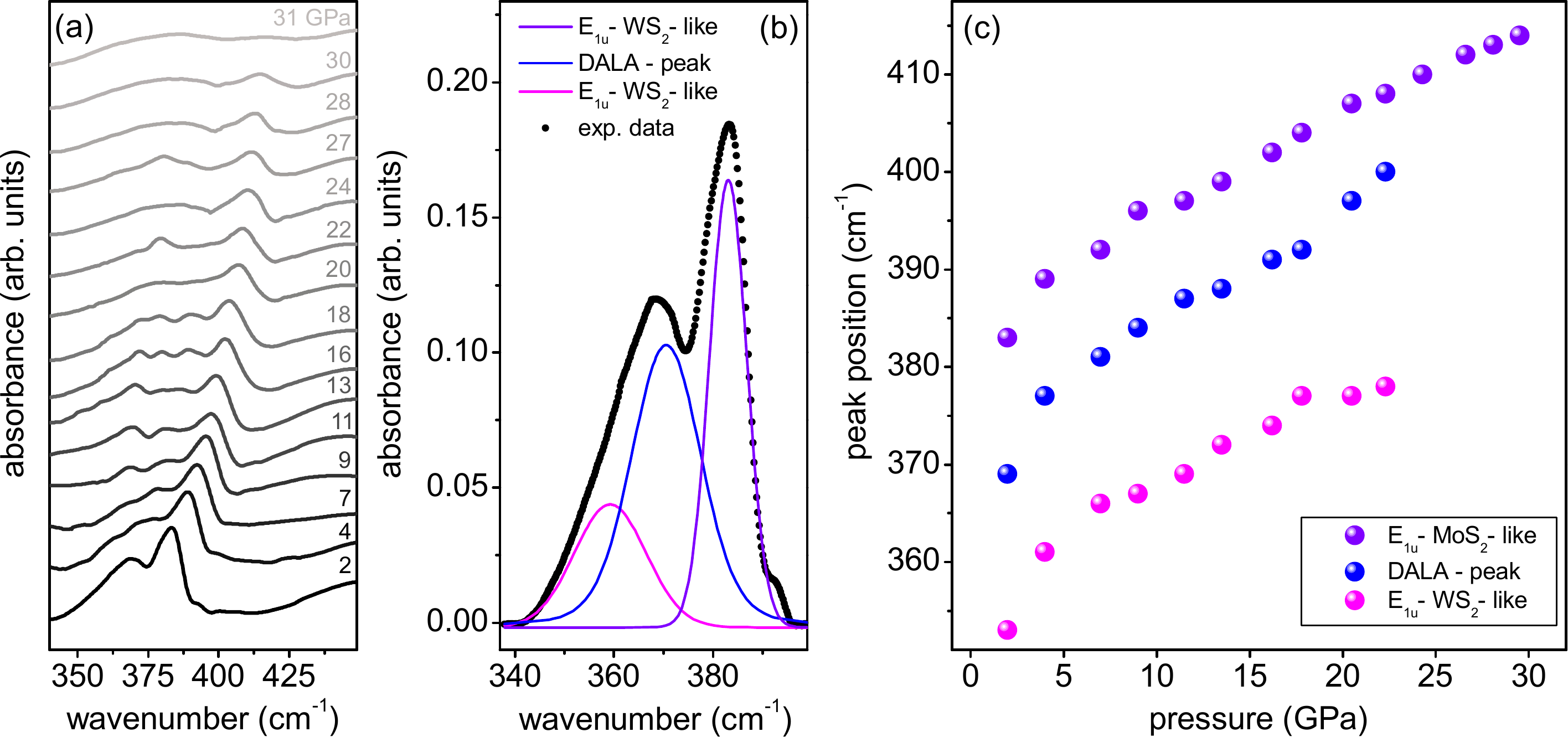}
  \caption{(a) Absorbance spectra of Mo$_{0.5}$W$_{0.5}$S$_2$ at increasing pressure from 2 to 31 GPa. Spectra are vertically shifted for sake of clarity. (b) Experimental phonon-peaks at 2 GPa (black dots) fitted with a sum of three Voigt functions. (c) Peak positions as a function of pressure. WS$_2$-like-E$_{1u}$ and DALA peaks are no longer distinguishable above 23 GPa.}
  \label{fig2}
\end{figure}

The absorbance spectra on increasing pressure in the 350$\div$430 cm$^{-1}$ range, where the phonon contribution is dominant, are shown in Figure \ref{fig2}. Although the overall intensity of the peaks is reduced with respect to the out-of-cell measurements, a comparison with the spectra reported in Figure \ref{fig1} allows us to identify the principal bands associated with E$_{1u}$-WS$_2$-like ($\sim$ 350 cm$^{-1}$), DALA ($\sim$ 370 cm$^{-1}$), and E$_{1u}$-MoS$_2$-like ($\sim$ 380 cm$^{-1}$) vibrational modes. In the cell measurements, the A$_{1g}$-like contributions, visible at ambient conditions at $\sim$ 410 cm$^{-1}$, are probably hindered by the interference fringes. Notice that, due to the alloyed nature of the compound, the relative intensity of the peaks may slightly vary depending on the measured point of the sample. In particular, disorder activated modes may be favoured in regions where the crystal order is reduced. By comparing Figures \ref{fig1} and \ref{fig2}, the relative intensity of the DALA mode is higher in the first case than in the second one. Since that the investigated samples and the regions of the sample surface are different for in-cell- and out-of-cell-measurements, this effect can be ascribed to a different impact of configurational disorder.
\\
As the pressure increases up to 31 GPa, the peak positions regularly shift toward higher frequencies, confirming the absence of structural transitions, as already suggested by Raman measurements \cite{Kim_2016}. The peak intensities above $\sim$ 20 GPa undergo a progressive lowering until completely vanishing at 31 GPa.

\subsection{Electronic properties at ambient conditions}
\begin{figure}
\includegraphics[scale=0.5]{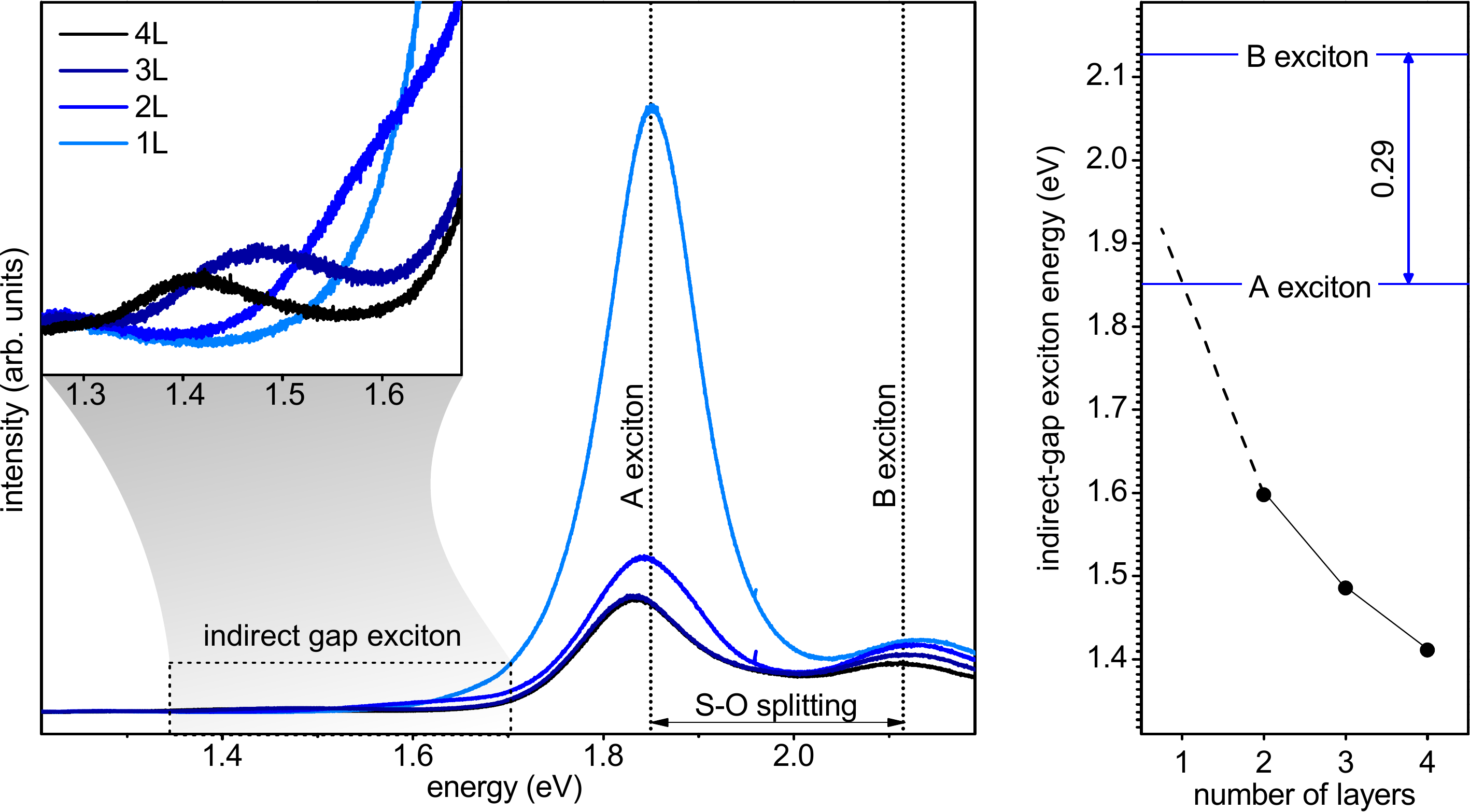}
  \caption{(a) PL spectrum of Mo$_{0.5}$W$_{0.5}$S$_2$ samples with a different number of layers. (b) Energy of the exciton associated with the indirect gap as a function of the number of layers.}
  \label{fig3}
\end{figure}

Thin Mo$_{0.5}$W$_{0.5}$S$_2$ crystals, mechanically exfoliated on a SiO$_2$-coated Si substrate, were analyzed through micro-Raman measurements to determine the number of layers (N) of each flake (see Supporting Information, SI). Photoluminescence (PL) measurements at ambient conditions were then performed to characterize the electronic properties of the sample as a function of N. PL spectra for mono-, bi-, tri- and four-layers samples are reported in Figure \ref{fig3}. The energy of the PL bands in the bulk sample is identical to that of the 4-layers sample, although the overall intensity is significantly lower. The spectrum of multi-layers samples exhibits three distinct bands: the one at lower energy ($\sim$ 1.4$\div$1.5 eV) corresponds to the exciton associated with the indirect-gap transition, the most intense one at $\sim$ 1.85 eV, is the A exciton, and the one at $\sim$ 2.1 eV is the B exciton, the latter two related to direct transitions \cite{Chen2013}. The energy difference between the peaks associated with A and B excitons corresponds to the spin-orbit splitting of the valence band of the crystal. In the mono-layer, the well-known transition from indirect to direct gap occurs, and the indirect transition band is no longer visible in the spectrum.
\\
A comparison between the PL spectra of Mo$_{0.5}$W$_{0.5}$S$_2$, MoS$_2$, and WS$_2$ (see SI) clearly shows that all the electronic features of the ternary compound lie at halfway between its basic constituents. In particular, in the bulk sample, we measured an indirect gap of $\sim$ 1.4 eV, a direct gap of $\sim$ 1.85 eV, and a spin-orbit splitting of $\sim$ 0.29 eV \cite{Chen2013}.

\subsection{Electronic properties under high-pressure}
\begin{figure}
\includegraphics[scale=0.5]{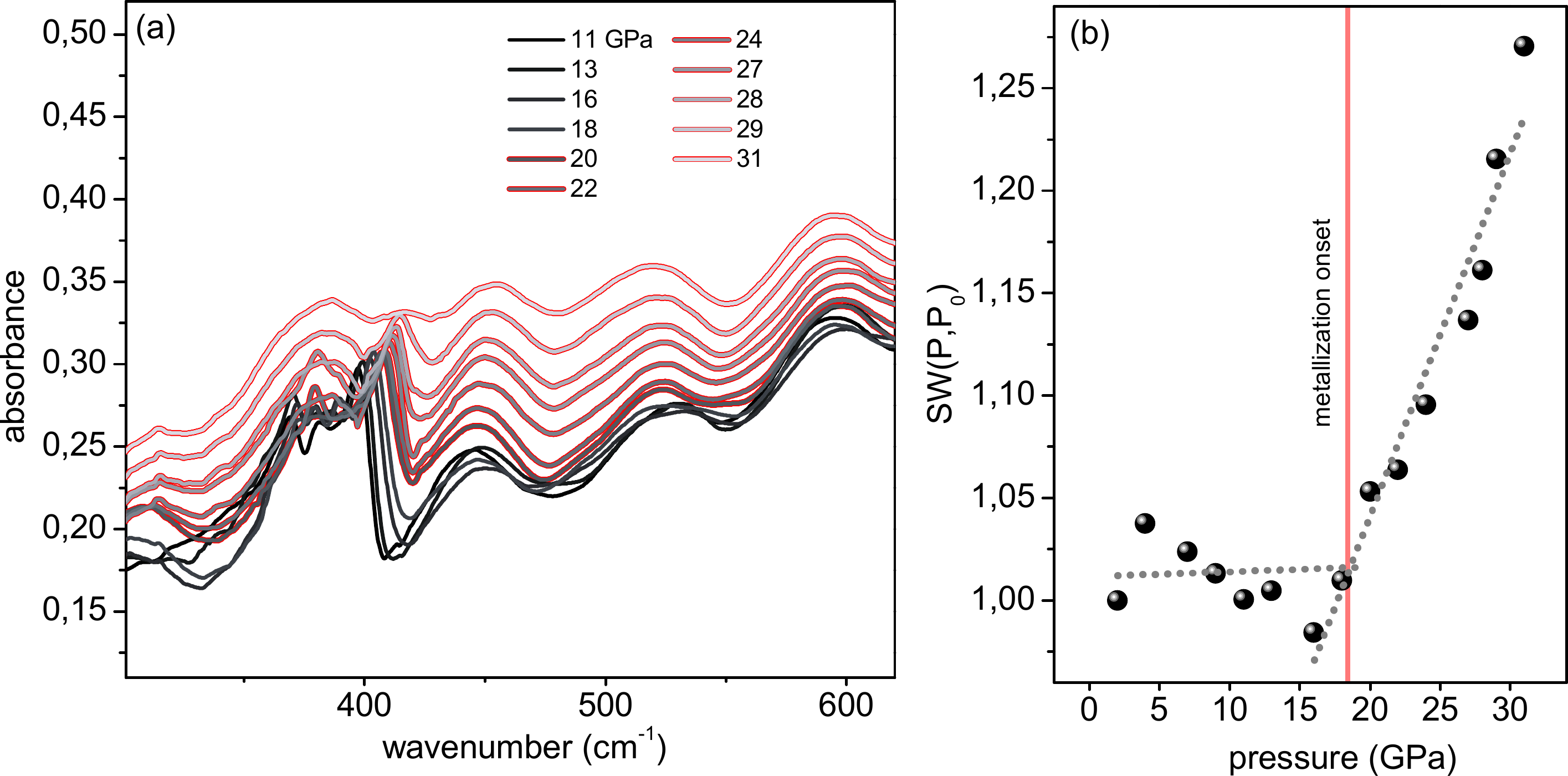}
  \caption{(a) Absorbance spectra of Mo$_{0.5}$W$_{0.5}$S$_2$ from 11 GPa to 31 GPa. Above 20 GPa, spectra with a red contour are characterized by increasing spectral weight. (b) Normalized spectral weight (SW(P,P$_0$)) as a function of pressure. Linear fits in the 2$\div$20 GPa and the 16$\div$31 GPa ranges are reported as dotted lines. }
  \label{fig4}
\end{figure}

To study the evolution of the electronic properties of Mo$_{0.5}$W$_{0.5}$S$_2$ as a function of pressure, we analyzed the spectral weight in the far-infrared range at increasing pressure values. This quantity indeed accounts for the low-energy electronic transitions and, thus, allows to monitor a possible insurgence of the metallic regime.
\\
As shown in Figure \ref{fig4}(a), the $A(\omega)$ spectra in the FIR region are almost superimposed up to $\sim$ 18 GPa. On further increasing pressure, the overall absorption intensity rises and the phonon peak intensity reduces. The first effect may arise from a rapid increase of the free electron density, due to the insurgence of a Drude band at zero frequency.
Indeed, as in most of semiconducting TMDs, in Mo$_{0.5}$W$_{0.5}$S$_2$ the high pressure reasonably drives a progressive reduction of the band-gap responsible for the semiconductor-to-metal transition. Correspondingly, the intensity of the vibrational modes decreases since the optical response of the free electrons shields the phonon contribution in the FIR absorption.
\\\\
To quantitatively characterize the metallization process, we define at each pressure the absorption spectral weight $\displaystyle{sw(P)=\int_{\omega_m}^{\omega_M}{A(\omega)d\omega}}$.  A simple application of the Drude model proves that in the low-frequency limit, i.e. $\omega/\Gamma \ll 1$  (where $\Gamma=c/\tau$ and $\tau$ is the Drude relaxation time), the absorbance is simply proportional to the square root of the DC conductivity $\sigma_0$ and $\omega$, $A(\omega) \propto \sqrt{\sigma_0 \omega}$. Since $\sigma_0$ is proportional to the carrier density n, the last relation demonstrates that an increase in the FIR absorbance is directly related to an increase of n. Figure \ref{fig4}(b) reports the values of the normalized spectral weight $SW(P,P_0)= sw(P)/sw(P_0)$ up to 31 GPa, where $P_0=2$ GPa and the integration limits are $\omega_m$=200 cm$^{-1}$ and $\omega_M$=700 cm$^{-1}$. The integration is not performed over the 350$\div$450 cm$^{-1}$ range where the phonon contribution is predominant. The normalized spectral weight, almost constant at low pressure, undergoes a significant increase around $\sim$ 20 GPa, associated to an increase of the charge carrier density n.  
To identify the onset of the metallization process, two independent linear fits have been performed in the 2$\div$20 GPa and the 16$\div$31 GPa regions, defining the metallization pressure, $P_M$, as the ordinate of the crossing point between the fitting lines: $P_{M} =18 \pm 1$ GPa, see Figure \ref{fig3}(b).
\\
The obtained results are well compatible with previous high-pressure measurements at room temperature on MoS$_2$ and WS$_2$, which have shown a metallic transition at  $\sim$ 20 GPa for both samples \cite{Nayak2014, Nayak2015}. Interestingly, in the Mo$_{0.5}$W$_{0.5}$S$_2$ alloy the substitutional disorder of the crystal lattice, attested by the presence of the disorder-activated phonon peaks, does not strongly affect the electronic properties of the sample.

\section{Experimental}
Mo$_{0.5}$W$_{0.5}$S$_2$, MoS$_2$, and WS$_2$ single crystals were provided by HQ-Graphene. All measurements were performed on fresh-cut, micrometric flakes, exfoliated from the macroscopic samples.
\\
Room temperature infrared transmission measurements were performed at the beamline SMIS of the SOLEIL synchrotron both at ambient pressure (on all the samples) and over the 0$\div$30 GPa range (only on Mo$_{0.5}$W$_{0.5}$S$_2$).
\\
In the DAC (orange), diamonds with a culet of 250 $\mu$m$^2$  were separated by a pre-indented stainless steel 50 $\mu$m thick gasket, in which a hole of 125 $\mu$m diameter was drilled. The exfoliated sample was positioned in the hole together with CsI as pressure transmitting medium \cite{Celeste2019} and a ruby chip, to measure the pressure through the ruby fluorescence technique \cite{Chijioke2005}. 
The measured sample was  50 $\mu$m$^2$ in area and 2-3 $\mu$m in thickness. 

Measurements were performed using a Thermo Fisher iS50 interferometer equipped with a solid-substrate beamsplitter. Synchrotron edge radiation was employed as light source.
Custom-made 15x Cassegrain objectives with a large working distance allowed to focus and then to collect the transmitted radiation, finally detected by a liquid-helium-cooled bolometer detector.

Raman and photoluminescence measurements at ambient conditions were collected on Mo$_{0.5}$W$_{0.5}$S$_2$, MoS$_2$ and WS$_2$, using a Horiba LabRAM HR Evolution microspectrometer, with a 532 nm laser as light source. The radiation was focused on a 2 $\mu$m$^2$ spot on the sample by a 100x objective and collected by a CCD detector. Further details are provided in Ref. \cite{caramazza2018first}.

\section{Conclusion}
Transmission measurements were performed on a Mo$_{0.5}$W$_{0.5}$S$_2$ single crystal in the far-infrared range. The spectrum collected at ambient conditions enabled the classification of the observed peaks in terms of the lattice vibrational modes, by comparison with the spectra of the binary constituents MoS$_2$ and WS$_2$. High-pressure measurements allowed us to study the evolution of both the lattice dynamics and the free carrier density up to 31 GPa. The absence of structural transitions, already suggested by Raman measurements, was confirmed, while an abrupt increase of the electron density was observed for the first time above 18 GPa. These results thus suggested the occurrence of an isostructural transition from a semiconducting to a metallic state, never investigated before in the literature. The observed effect is coherent with previous studies on the binary compounds MoS$_2$ and WS$_2$, which reported the onset of the metallization process at $\sim$ 19 GPa \cite{Nayak2014} and $\sim$ 22 GPa \cite{Nayak2015}, respectively, without relevant changes in the lattice symmetries. The compatibility between the metallization pressures of Mo$_{0.5}$W$_{0.5}$S$_2$ and its constituents suggests that substitutional disorder does not play a key role in the evolution of the electronic properties of the alloy under pressure.

\bibliography{achemso-demo}

\providecommand{\latin}[1]{#1}
\makeatletter
\providecommand{\doi}
  {\begingroup\let\do\@makeother\dospecials
  \catcode`\{=1 \catcode`\}=2 \doi@aux}
\providecommand{\doi@aux}[1]{\endgroup\texttt{#1}}
\makeatother
\providecommand*\mcitethebibliography{\thebibliography}
\csname @ifundefined\endcsname{endmcitethebibliography}
  {\let\endmcitethebibliography\endthebibliography}{}
\begin{mcitethebibliography}{33}
\providecommand*\natexlab[1]{#1}
\providecommand*\mciteSetBstSublistMode[1]{}
\providecommand*\mciteSetBstMaxWidthForm[2]{}
\providecommand*\mciteBstWouldAddEndPuncttrue
  {\def\EndOfBibitem{\unskip.}}
\providecommand*\mciteBstWouldAddEndPunctfalse
  {\let\EndOfBibitem\relax}
\providecommand*\mciteSetBstMidEndSepPunct[3]{}
\providecommand*\mciteSetBstSublistLabelBeginEnd[3]{}
\providecommand*\EndOfBibitem{}
\mciteSetBstSublistMode{f}
\mciteSetBstMaxWidthForm{subitem}{(\alph{mcitesubitemcount})}
\mciteSetBstSublistLabelBeginEnd
  {\mcitemaxwidthsubitemform\space}
  {\relax}
  {\relax}

\bibitem[Yuan \latin{et~al.}(2016)Yuan, Ge, Peng, Zhang, Wu, Jian, Xiong, Yin,
  and Han]{Yuan2016}
Yuan,~L.; Ge,~J.; Peng,~X.; Zhang,~Q.; Wu,~Z.; Jian,~Y.; Xiong,~X.; Yin,~H.;
  Han,~J. A reliable way of mechanical exfoliation of large scale two
  dimensional materials with high quality. \emph{AIP Adv.} \textbf{2016},
  \emph{6}, 125201\relax
\mciteBstWouldAddEndPuncttrue
\mciteSetBstMidEndSepPunct{\mcitedefaultmidpunct}
{\mcitedefaultendpunct}{\mcitedefaultseppunct}\relax
\EndOfBibitem
\bibitem[Manzeli \latin{et~al.}(2017)Manzeli, Ovchinnikov, Pasquier, Yazyev,
  and Kis]{Manzeli2017}
Manzeli,~S.; Ovchinnikov,~D.; Pasquier,~D.; Yazyev,~O.~V.; Kis,~A. 2D
  transition metal dichalcogenides. \emph{Nature Reviews Materials}
  \textbf{2017}, \emph{2}\relax
\mciteBstWouldAddEndPuncttrue
\mciteSetBstMidEndSepPunct{\mcitedefaultmidpunct}
{\mcitedefaultendpunct}{\mcitedefaultseppunct}\relax
\EndOfBibitem
\bibitem[Peng \latin{et~al.}(2017)Peng, Wu, Li, Zhou, Yu, Guo, Wu, Lin, Li, Wu,
  Wu, and Xie]{Peng2017}
Peng,~J.; Wu,~J.; Li,~X.; Zhou,~Y.; Yu,~Z.; Guo,~Y.; Wu,~J.; Lin,~Y.; Li,~Z.;
  Wu,~X.; Wu,~C.; Xie,~Y. Very Large-Sized Transition Metal Dichalcogenides
  Monolayers from Fast Exfoliation by Manual Shaking. \emph{Journal of the
  American Chemical Society} \textbf{2017}, \emph{139}, 9019--9025\relax
\mciteBstWouldAddEndPuncttrue
\mciteSetBstMidEndSepPunct{\mcitedefaultmidpunct}
{\mcitedefaultendpunct}{\mcitedefaultseppunct}\relax
\EndOfBibitem
\bibitem[Li \latin{et~al.}(2017)Li, Tao, Chen, Fang, Li, Wang, Xu, and
  Zhu]{Li2017}
Li,~X.; Tao,~L.; Chen,~Z.; Fang,~H.; Li,~X.; Wang,~X.; Xu,~J.~B.; Zhu,~H.
  Graphene and related two-dimensional materials: Structure-property
  relationships for electronics and optoelectronics. \emph{App. Phys. Rev.}
  \textbf{2017}, \emph{4}, 021306\relax
\mciteBstWouldAddEndPuncttrue
\mciteSetBstMidEndSepPunct{\mcitedefaultmidpunct}
{\mcitedefaultendpunct}{\mcitedefaultseppunct}\relax
\EndOfBibitem
\bibitem[Wang \latin{et~al.}(2012)Wang, Kalantar-Zadeh, Kis, Coleman, and
  Strano]{Wang2012}
Wang,~Q.~H.; Kalantar-Zadeh,~K.; Kis,~A.; Coleman,~J.~N.; Strano,~M.~S.
  Electronics and optoelectronics of two-dimensional transition metal
  dichalcogenides. \emph{Nature Nanotechnology} \textbf{2012}, \emph{7},
  699--712\relax
\mciteBstWouldAddEndPuncttrue
\mciteSetBstMidEndSepPunct{\mcitedefaultmidpunct}
{\mcitedefaultendpunct}{\mcitedefaultseppunct}\relax
\EndOfBibitem
\bibitem[Jariwala \latin{et~al.}(2014)Jariwala, Sangwan, Lauhon, Marks, and
  Hersam]{Jariwala2014}
Jariwala,~D.; Sangwan,~V.~K.; Lauhon,~L.~J.; Marks,~T.~J.; Hersam,~M.~C.
  Emerging Device Applications for Semiconducting Two-Dimensional Transition
  Metal Dichalcogenides. \emph{{ACS} Nano} \textbf{2014}, \emph{8},
  1102--1120\relax
\mciteBstWouldAddEndPuncttrue
\mciteSetBstMidEndSepPunct{\mcitedefaultmidpunct}
{\mcitedefaultendpunct}{\mcitedefaultseppunct}\relax
\EndOfBibitem
\bibitem[Yun \latin{et~al.}(2012)Yun, Han, Hong, and Lee]{Yun2012}
Yun,~W.~S.; Han,~S.~W.; Hong,~I.~G.,~S. C.and~Kim; Lee,~J.~D. Thickness and
  strain effects on electronic structures of transition metal dichalcogenides:
  2H-MX$_2$ semiconductors (M = Mo, W; X = S, Se, Te). \emph{Phys. Rev. B}
  \textbf{2012}, \emph{85}, 033305\relax
\mciteBstWouldAddEndPuncttrue
\mciteSetBstMidEndSepPunct{\mcitedefaultmidpunct}
{\mcitedefaultendpunct}{\mcitedefaultseppunct}\relax
\EndOfBibitem
\bibitem[Mak \latin{et~al.}(2010)Mak, Lee, Hone, Shan, and Heinz]{Mak2010}
Mak,~K.~F.; Lee,~C.; Hone,~J.; Shan,~J.; Heinz,~T.~F. Atomically Thin MoS$_2$:
  A New Direct-Gap Semiconductor. \emph{Phys. Rev. Lett.} \textbf{2010},
  \emph{105}, 136805\relax
\mciteBstWouldAddEndPuncttrue
\mciteSetBstMidEndSepPunct{\mcitedefaultmidpunct}
{\mcitedefaultendpunct}{\mcitedefaultseppunct}\relax
\EndOfBibitem
\bibitem[y.~Zhang~et al.(2013)]{Zhang2013}
y.~Zhang~et al., Direct observation of the transition from indirect to direct
  bandgap in atomically thin epitaxial MoSe$_2$. \emph{Nat. Nanotechnol.}
  \textbf{2013}, \emph{9}, 111 EP --\relax
\mciteBstWouldAddEndPuncttrue
\mciteSetBstMidEndSepPunct{\mcitedefaultmidpunct}
{\mcitedefaultendpunct}{\mcitedefaultseppunct}\relax
\EndOfBibitem
\bibitem[Xie(2015)]{Xie2015}
Xie,~L.~M. Two-dimensional transition metal dichalcogenide alloys:
  preparation{,} characterization and applications. \emph{Nanoscale}
  \textbf{2015}, \emph{7}, 18392--18401\relax
\mciteBstWouldAddEndPuncttrue
\mciteSetBstMidEndSepPunct{\mcitedefaultmidpunct}
{\mcitedefaultendpunct}{\mcitedefaultseppunct}\relax
\EndOfBibitem
\bibitem[Zhao \latin{et~al.}(2018)Zhao, Zhang, and Ouyang]{Zhao2018}
Zhao,~Y.; Zhang,~Z.; Ouyang,~G. Band shift of 2D transition-metal
  dichalcogenide alloys: size and composition effects. \emph{Applied Physics A}
  \textbf{2018}, \emph{124}\relax
\mciteBstWouldAddEndPuncttrue
\mciteSetBstMidEndSepPunct{\mcitedefaultmidpunct}
{\mcitedefaultendpunct}{\mcitedefaultseppunct}\relax
\EndOfBibitem
\bibitem[Sahoo \latin{et~al.}(2013)Sahoo, Gaur, Ahmadi, Guinel, and
  Katiyar]{Sahoo2013}
Sahoo,~S.; Gaur,~A. P.~S.; Ahmadi,~M.; Guinel,~M. J.-F.; Katiyar,~R.~S.
  Temperature-Dependent Raman Studies and Thermal Conductivity of Few-Layer
  {MoS}2. \emph{The Journal of Physical Chemistry C} \textbf{2013}, \emph{117},
  9042--9047\relax
\mciteBstWouldAddEndPuncttrue
\mciteSetBstMidEndSepPunct{\mcitedefaultmidpunct}
{\mcitedefaultendpunct}{\mcitedefaultseppunct}\relax
\EndOfBibitem
\bibitem[Chakraborty \latin{et~al.}(2012)Chakraborty, Matte, Sood, and
  Rao]{Chakraborty2012}
Chakraborty,~B.; Matte,~H. S. S.~R.; Sood,~A.~K.; Rao,~C. N.~R. Layer-dependent
  resonant Raman scattering of a few layer {MoS}2. \emph{Journal of Raman
  Spectroscopy} \textbf{2012}, \emph{44}, 92--96\relax
\mciteBstWouldAddEndPuncttrue
\mciteSetBstMidEndSepPunct{\mcitedefaultmidpunct}
{\mcitedefaultendpunct}{\mcitedefaultseppunct}\relax
\EndOfBibitem
\bibitem[Mahatha \latin{et~al.}(2012)Mahatha, Patel, and Menon]{Mahatha_2012}
Mahatha,~S.~K.; Patel,~K.~D.; Menon,~K. S.~R. Electronic structure
  investigation of {MoS}2and {MoSe}2using angle-resolved photoemission
  spectroscopy andab initioband structure studies. \emph{Journal of Physics:
  Condensed Matter} \textbf{2012}, \emph{24}, 475504\relax
\mciteBstWouldAddEndPuncttrue
\mciteSetBstMidEndSepPunct{\mcitedefaultmidpunct}
{\mcitedefaultendpunct}{\mcitedefaultseppunct}\relax
\EndOfBibitem
\bibitem[Ulstrup \latin{et~al.}(2017)Ulstrup, \ifmmode~\check{C}\else
  \v{C}\fi{}abo, Biswas, Riley, Dendzik, Sanders, Bianchi, Cacho, Matselyukh,
  Chapman, Springate, King, Miwa, and Hofmann]{Ulstrup2017}
Ulstrup,~S.; \ifmmode~\check{C}\else \v{C}\fi{}abo,~A. G. c. v. a.~c.;
  Biswas,~D.; Riley,~J.~M.; Dendzik,~M.; Sanders,~C.~E.; Bianchi,~M.;
  Cacho,~C.; Matselyukh,~D.; Chapman,~R.~T.; Springate,~E.; King,~P. D.~C.;
  Miwa,~J.~A.; Hofmann,~P. Spin and valley control of free carriers in
  single-layer ${\mathrm{WS}}_{2}$. \emph{Phys. Rev. B} \textbf{2017},
  \emph{95}, 041405\relax
\mciteBstWouldAddEndPuncttrue
\mciteSetBstMidEndSepPunct{\mcitedefaultmidpunct}
{\mcitedefaultendpunct}{\mcitedefaultseppunct}\relax
\EndOfBibitem
\bibitem[Sekine \latin{et~al.}(1980)Sekine, Nakashizu, Toyoda, Uchinokura, and
  Matsuura]{Sekine1980}
Sekine,~T.; Nakashizu,~T.; Toyoda,~K.; Uchinokura,~K.; Matsuura,~E. Raman
  scattering in layered compound 2H-{WS}2. \emph{Solid State Communications}
  \textbf{1980}, \emph{35}, 371--373\relax
\mciteBstWouldAddEndPuncttrue
\mciteSetBstMidEndSepPunct{\mcitedefaultmidpunct}
{\mcitedefaultendpunct}{\mcitedefaultseppunct}\relax
\EndOfBibitem
\bibitem[Wang \latin{et~al.}(2016)Wang, Liu, Ito, Ning, Tan, Fujita, Hirata,
  and Chen]{Wang2016}
Wang,~Z.; Liu,~P.; Ito,~Y.; Ning,~S.; Tan,~Y.; Fujita,~T.; Hirata,~A.; Chen,~M.
  Chemical Vapor Deposition of Monolayer Mo1-{xWxS}2 Crystals with Tunable Band
  Gaps. \emph{Scientific Reports} \textbf{2016}, \emph{6}\relax
\mciteBstWouldAddEndPuncttrue
\mciteSetBstMidEndSepPunct{\mcitedefaultmidpunct}
{\mcitedefaultendpunct}{\mcitedefaultseppunct}\relax
\EndOfBibitem
\bibitem[Kim \latin{et~al.}(2017)Kim, Ahmad, Pandey, Rai, Feng, Yang, Lin,
  Terrones, Banerjee, Singh, Akinwande, and Lin]{Kim2017}
Kim,~J.-S.; Ahmad,~R.; Pandey,~T.; Rai,~A.; Feng,~S.; Yang,~J.; Lin,~Z.;
  Terrones,~M.; Banerjee,~S.~K.; Singh,~A.~K.; Akinwande,~D.; Lin,~J.-F.
  Towards band structure and band offset engineering of monolayer
  Mo$_x$W$_{1-x}$S$_2$ via Strain. \emph{2D Materials} \textbf{2017}, \emph{5},
  015008\relax
\mciteBstWouldAddEndPuncttrue
\mciteSetBstMidEndSepPunct{\mcitedefaultmidpunct}
{\mcitedefaultendpunct}{\mcitedefaultseppunct}\relax
\EndOfBibitem
\bibitem[Qiao \latin{et~al.}(2015)Qiao, Li, Zhang, Shi, Wu, Chen, and
  Tan]{Qiao2015}
Qiao,~X.-F.; Li,~X.-L.; Zhang,~X.; Shi,~W.; Wu,~J.-B.; Chen,~T.; Tan,~P.-H.
  Substrate-free layer-number identification of two-dimensional materials: A
  case of Mo0.5W0.5S2 alloy. \emph{Applied Physics Letters} \textbf{2015},
  \emph{106}, 223102\relax
\mciteBstWouldAddEndPuncttrue
\mciteSetBstMidEndSepPunct{\mcitedefaultmidpunct}
{\mcitedefaultendpunct}{\mcitedefaultseppunct}\relax
\EndOfBibitem
\bibitem[Zhang \latin{et~al.}(2015)Zhang, Qiao, Shi, Wu, Jiang, and
  Tan]{Zhang2015}
Zhang,~X.; Qiao,~X.-F.; Shi,~W.; Wu,~J.-B.; Jiang,~D.-S.; Tan,~P.-H. Phonon and
  Raman scattering of two-dimensional transition metal dichalcogenides from
  monolayer, multilayer to bulk material. \emph{Chemical Society Reviews}
  \textbf{2015}, \emph{44}, 2757--2785\relax
\mciteBstWouldAddEndPuncttrue
\mciteSetBstMidEndSepPunct{\mcitedefaultmidpunct}
{\mcitedefaultendpunct}{\mcitedefaultseppunct}\relax
\EndOfBibitem
\bibitem[Stellino(2020)]{Stellino2020}
Stellino,~E. Pressure evolution of the optical phonons of MoTe2. \emph{Il Nuovo
  Cimento C} \textbf{2020}, \emph{43}, 1–8\relax
\mciteBstWouldAddEndPuncttrue
\mciteSetBstMidEndSepPunct{\mcitedefaultmidpunct}
{\mcitedefaultendpunct}{\mcitedefaultseppunct}\relax
\EndOfBibitem
\bibitem[Yang \latin{et~al.}(2019)Yang, Dai, Li, Hu, Liu, Pu, Hong, and
  Liu]{Yang2019}
Yang,~L.; Dai,~L.; Li,~H.; Hu,~H.; Liu,~K.; Pu,~C.; Hong,~M.; Liu,~P.
  Pressure-induced metallization in {MoSe}2 under different pressure
  conditions. \emph{{RSC} Advances} \textbf{2019}, \emph{9}, 5794--5803\relax
\mciteBstWouldAddEndPuncttrue
\mciteSetBstMidEndSepPunct{\mcitedefaultmidpunct}
{\mcitedefaultendpunct}{\mcitedefaultseppunct}\relax
\EndOfBibitem
\bibitem[Nayak \latin{et~al.}(2014)Nayak, Bhattacharyya, Zhu, Liu, Wu, Pandey,
  Jin, Singh, Akinwande, and Lin]{Nayak2014}
Nayak,~A.~P.; Bhattacharyya,~S.; Zhu,~J.; Liu,~J.; Wu,~X.; Pandey,~T.; Jin,~C.;
  Singh,~A.~K.; Akinwande,~D.; Lin,~J.-F. Pressure-induced semiconducting to
  metallic transition in multilayered molybdenum disulphide. \emph{Nature
  Communications} \textbf{2014}, \emph{5}\relax
\mciteBstWouldAddEndPuncttrue
\mciteSetBstMidEndSepPunct{\mcitedefaultmidpunct}
{\mcitedefaultendpunct}{\mcitedefaultseppunct}\relax
\EndOfBibitem
\bibitem[Nayak \latin{et~al.}(2015)Nayak, Yuan, Cao, Liu, Wu, Moran, Li,
  Akinwande, Jin, and Lin]{Nayak2015}
Nayak,~A.~P.; Yuan,~Z.; Cao,~B.; Liu,~J.; Wu,~J.; Moran,~S.~T.; Li,~T.;
  Akinwande,~D.; Jin,~C.; Lin,~J.-F. Pressure-Modulated Conductivity, Carrier
  Density, and Mobility of Multilayered Tungsten Disulfide. \emph{{ACS} Nano}
  \textbf{2015}, \emph{9}, 9117--9123\relax
\mciteBstWouldAddEndPuncttrue
\mciteSetBstMidEndSepPunct{\mcitedefaultmidpunct}
{\mcitedefaultendpunct}{\mcitedefaultseppunct}\relax
\EndOfBibitem
\bibitem[Kim \latin{et~al.}(2016)Kim, Moran, Nayak, Pedahzur, Ruiz, Ponce,
  Rodriguez, Henny, Liu, Lin, and Akinwande]{Kim_2016}
Kim,~J.-S.; Moran,~S.~T.; Nayak,~A.~P.; Pedahzur,~S.; Ruiz,~I.; Ponce,~G.;
  Rodriguez,~D.; Henny,~J.; Liu,~J.; Lin,~J.-F.; Akinwande,~D. High pressure
  Raman study of layered Mo$_{0.5}$W$_{0.5}$S$_2$ ternary compound. \emph{2D
  Materials} \textbf{2016}, \emph{3}, 025003\relax
\mciteBstWouldAddEndPuncttrue
\mciteSetBstMidEndSepPunct{\mcitedefaultmidpunct}
{\mcitedefaultendpunct}{\mcitedefaultseppunct}\relax
\EndOfBibitem
\bibitem[O'Neal \latin{et~al.}(2016)O'Neal, Cherian, Zak, Tenne, Liu, and
  Musfeldt]{ONeal2016}
O'Neal,~K.~R.; Cherian,~J.~G.; Zak,~A.; Tenne,~R.; Liu,~Z.; Musfeldt,~J.~L.
  High Pressure Vibrational Properties of {WS}2 Nanotubes. \emph{Nano Letters}
  \textbf{2016}, \emph{16}, 993--999\relax
\mciteBstWouldAddEndPuncttrue
\mciteSetBstMidEndSepPunct{\mcitedefaultmidpunct}
{\mcitedefaultendpunct}{\mcitedefaultseppunct}\relax
\EndOfBibitem
\bibitem[Guo \latin{et~al.}(2015)Guo, Chen, Wen, and Zheng]{Guo2015}
Guo,~X.; Chen,~H.; Wen,~X.; Zheng,~J. Electron-phonon interactions in {MoS}2
  probed with ultrafast two-dimensional visible/far-infrared spectroscopy.
  \emph{The Journal of Chemical Physics} \textbf{2015}, \emph{142},
  212447\relax
\mciteBstWouldAddEndPuncttrue
\mciteSetBstMidEndSepPunct{\mcitedefaultmidpunct}
{\mcitedefaultendpunct}{\mcitedefaultseppunct}\relax
\EndOfBibitem
\bibitem[Pike \latin{et~al.}(2017)Pike, Van~Troeye, Dewandre, Petretto, Gonze,
  Rignanese, and Verstraete]{Pike2017}
Pike,~N.~A.; Van~Troeye,~B.; Dewandre,~A.; Petretto,~G.; Gonze,~X.;
  Rignanese,~G.-M.; Verstraete,~M.~J. Origin of the counterintuitive dynamic
  charge in the transition metal dichalcogenides. \emph{Phys. Rev. B}
  \textbf{2017}, \emph{95}, 201106\relax
\mciteBstWouldAddEndPuncttrue
\mciteSetBstMidEndSepPunct{\mcitedefaultmidpunct}
{\mcitedefaultendpunct}{\mcitedefaultseppunct}\relax
\EndOfBibitem
\bibitem[Chen \latin{et~al.}(2013)Chen, Xi, Dumcenco, Liu, Suenaga, Wang,
  Shuai, Huang, and Xie]{Chen2013}
Chen,~Y.; Xi,~J.; Dumcenco,~D.~O.; Liu,~Z.; Suenaga,~K.; Wang,~D.; Shuai,~Z.;
  Huang,~Y.-S.; Xie,~L. Tunable Band Gap Photoluminescence from Atomically Thin
  Transition-Metal Dichalcogenide Alloys. \emph{{ACS} Nano} \textbf{2013},
  \emph{7}, 4610--4616\relax
\mciteBstWouldAddEndPuncttrue
\mciteSetBstMidEndSepPunct{\mcitedefaultmidpunct}
{\mcitedefaultendpunct}{\mcitedefaultseppunct}\relax
\EndOfBibitem
\bibitem[Celeste \latin{et~al.}(2019)Celeste, Borondics, and
  Capitani]{Celeste2019}
Celeste,~A.; Borondics,~F.; Capitani,~F. Hydrostaticity of
  pressure-transmitting media for high pressure infrared spectroscopy.
  \emph{High Pressure Research} \textbf{2019}, \emph{39}, 608--618\relax
\mciteBstWouldAddEndPuncttrue
\mciteSetBstMidEndSepPunct{\mcitedefaultmidpunct}
{\mcitedefaultendpunct}{\mcitedefaultseppunct}\relax
\EndOfBibitem
\bibitem[Chijioke \latin{et~al.}(2005)Chijioke, Nellis, Soldatov, and
  Silvera]{Chijioke2005}
Chijioke,~A.~D.; Nellis,~W.~J.; Soldatov,~A.; Silvera,~I.~F. The ruby pressure
  standard to 150GPa. \emph{Journal of Applied Physics} \textbf{2005},
  \emph{98}, 114905\relax
\mciteBstWouldAddEndPuncttrue
\mciteSetBstMidEndSepPunct{\mcitedefaultmidpunct}
{\mcitedefaultendpunct}{\mcitedefaultseppunct}\relax
\EndOfBibitem
\bibitem[Caramazza \latin{et~al.}(2018)Caramazza, Collina, Stellino, Ripanti,
  Dore, and Postorino]{caramazza2018first}
Caramazza,~S.; Collina,~A.; Stellino,~E.; Ripanti,~F.; Dore,~P.; Postorino,~P.
  First-and second-order Raman scattering from MoTe 2 single crystal. \emph{The
  European Physical Journal B} \textbf{2018}, \emph{91}, 1--7\relax
\mciteBstWouldAddEndPuncttrue
\mciteSetBstMidEndSepPunct{\mcitedefaultmidpunct}
{\mcitedefaultendpunct}{\mcitedefaultseppunct}\relax
\EndOfBibitem
\end{mcitethebibliography}

\end{document}